\documentclass[amsmath,amssymb,aps,pra,reprint,groupedaddress]{revtex4-1}

\usepackage{graphicx}
\usepackage{dcolumn}
\usepackage{bm}

\begin{document}


\title{Leveraging beam deformation to improve the detection of resonances}


\author{R\'emi Poll\`es}
\author{Martine Mihaïlovic}
\author{Emmanuel Centeno}
\author{Antoine Moreau}
\affiliation{Universit\'e Clermont Auvergne, Institut Pascal, PHOTON-N2, BP 10448, F-63000 CLERMONT-FERRAND, FRANCE}
\affiliation{CNRS, UMR 6602, IP, F-63178 Aubi\`ere, FRANCE}

\date{\today}

\begin{abstract}
Decades of work on beam deformation on reflection, and especially on lateral shifts, have spread the idea that a reflected beam is larger than the incident beam. However, when the right conditions are met, a beam reflected by a multilayered resonant structure can be 10\% narrower than the incoming beam. Such an easily measurable change occurs on a very narrow angular range close to a resonance, which can be leveraged to improve the resolution of sensors based on the detection of surface plasmon resonances by a factor three.  We provide theoretical tools to deal with this effect, and a thorough physical discussion that leads to expect similar phenomenon to occur for temporal wavepackets and in other domains of physics. 
\end{abstract}

\maketitle

Detecting an optical resonance classically reduces to sending a beam on a structure and using the amount of light that is reflected (or transmitted) to accurately determine for which angle or for which frequency the resonance occurs. The beam is almost always large enough so that the finite size of the beam has no influence on the measurement. On the other hand, it is common knowledge that when a narrow enough beam is reflected by a bare interface, the reflected beam can be deformed by the reflection to the extent that even the basic laws of the specular reflection do not seem to hold any more\cite{tamir86,merano}. The changes are thus said to be {\em nonspecular} and have been continuously explored since Newton\cite{newton,lotsch}. Especially, the lateral shift of reflected beams has attracted most of the attention\cite{resonantgh,nonlineargh,felbacq03,li03,yin06,hosten08,bliokh08,wang13}, after the pioneering experimental work of Goos and H\"anchen and some theoretical work in the seventies\cite{tamir71a,tamir71b}. The study of lateral shifts, especially large ones which result from the excitation of leaky modes, leads to the conclusion that, in general, the reflected beam is always larger than the incoming one. This corresponds to the commonly shared idea that, in physics, the deformation of a wavepacket by a linear physical phenomenon leads to a widening and to dispersion. For these reasons, very little attention has ever been paid to the change in width undergone by a beam when it is reflected by a multilayered structure. 

Here we show that, for any multilayered structure whose resonance leads to a reflection dip, the reflected beam can actually be narrower than the incoming beam because of a destructive interference between the beam reflected on the first interface and the resonance. This phenomenon can be leveraged to push the theoretical resolution limits of Surface Plasmon Resonance (SPR) detection\cite{piliarik09}, as it occurs on an angular range that is narrower (typically three times) than the range on which the reflection coefficient varies. We provide analytical formulae to describe the  variation in the beam width on reflection and a thorough physical analysis why this change occurs on such a narrow angular range. Conversely to many nonspecular phenomena that have been predicted relying on formulas that are valid only for very large beams\cite{artmann}, hindering their use for any pratical application, the phenomenon we want to monitor occurs for finite, realistic beams. Finally, we underline that the validity of our analysis extends to temporal wavepackets and other domains of physics, as electronics and quantum mechanics.



We introduce first the very general formulas that describe how a beam is shifted and widened or narrowed on reflection on a multilayered structure {\em whatever the width of the incoming beam}, and not just in large waist limit that most of the authors, following Artmann\cite{artmann}, consider.

The electric field, in $s$ polarization, of an incoming beam can be described in terms of its plane wave expansion, each plane wave being characterized by a wavevector $\alpha = n \,k_0\,\sin \theta$ where $\theta$ is the associated incidence angle and $n$ the optical index of the medium. It can thus be written
\begin{equation}
E_i(x,z,\omega)=\frac{1}{2\pi}\int \tilde{E_i} \left(\alpha\right) e^{i\left(\alpha x -\gamma z -\omega t \right)}d\alpha,
\end{equation}
where $\gamma=\sqrt{\varepsilon\, \mu\,k_0^2-\alpha^2}$ and where the angular spectral amplitude is given by 
\begin{equation}
\tilde{E_i} \left(\alpha \right) = \frac{w}{2 \sqrt{\pi}} e^{-\frac{w_i^2}{4} \left(\alpha-\alpha_0 \right)^2}.
\label{eq:gaussian}
\end{equation}
This corresponds to a gaussian beam with a waist $w_i$, angularly centered on $\theta_0$ with $\alpha_0=n\, k_0\,\sin\,\theta_0$
where $k_0=\frac{2\pi}{\lambda}$. Light is reflected by the structure beginning at $z=0$, producing a beam with an angular spectrum 
\begin{equation}
\tilde{E_r}=r(\alpha) \tilde{E_i}= \rho(\alpha) \,e^{i\phi(\alpha)}\,\tilde{E_i}.
\end{equation}
where $r$ is the reflection coefficient, and $\rho$ and $\phi$ its modulus and phase, respectively. This formalism of course holds to describe the $H_y$ field in $p$-polarization.

The lateral shift on reflection is the difference between the center of the reflected beam and the center of the incoming beam, 
\begin{equation}\label{delta}
\delta=\frac{\int x |E_r|^2 dx}{\int |E_r|^2 dx}-\frac{\int x |E_i|^2 dx}{\int |E_i|^2 dx}.
\end{equation}
It is possible to show (see Appendix A) that this shift is given {\em whatever the beam width}, and even if the modulus of the reflection coefficient is not $1$ by 
\begin{equation}
\delta=-\frac{\int \rho^2 \phi' |\tilde{E_i}|^2 d\alpha}{\int \rho^2 |\tilde{E_i}|^2 d\alpha}.
\end{equation}
where $'$ denotes a derivation with respect to $\alpha$.

We are interested here in the change in the width of the beam on reflection\cite{tamir86}. It can be simply defined as a difference between second-order centered moments, just like the shift is a difference between first-order moments : 
\begin{equation}
\Delta=\frac{\int (x-\delta)^2 |E_r|^2 dx}{\int |E_r|^2 dx}-\frac{\int x^2 |E_i|^2 dx}{\int |E_i|^2 dx}.
\end{equation}
We underline that, with the above definition of a Gaussian beam, we have $\int x^2 |E_i|^2 dx/\int |E_i|^2 dx = \frac{1}{4}\,w_i^2$, clearly showing how meaningful the second order moments are. If the reflected beam can be considered gaussian with a waist $w_r$, then we have $\Delta = \frac{1}{4} (w_r^2-w_i^2)$, but usually the reflected beam is not rigorously gaussian. Using the same kind of demonstration as for the shift, a relatively straightforward calculation (see Appendix B) yields
\begin{multline}
\Delta = \frac{ \int\frac{1}{2}(\rho'^2-\rho\rho'')|\tilde{E_i}|^2 d\alpha } { \int \rho^2|\tilde{E_i}|^2 d\alpha }
+\left[\frac{ \int\rho^2\frac{w^2}{4}\tilde{E_i}^2 d\alpha } { \int \rho^2|\tilde{E_i}|^2 d\alpha }\right.\\
\left. - \frac{ \int\frac{w^2}{4}\tilde{E_i}^2 d\alpha } { \int |\tilde{E_i}|^2 d\alpha }\right]
+\left[\frac{ \int \rho^2\phi'^2|\tilde{E_i}|^2 d\alpha } { \int \rho^2|\tilde{E_i}|^2 d\alpha }-\delta^2\right].
\label{eq:expansion1}
\end{multline}

When the width of the incoming beam becomes asymptotically large, its angular spectrum becomes a Dirac distribution, so that the lateral shift tends to a finite limit $\delta\rightarrow -\phi'$. This is Artmann's\cite{artmann} formula

\begin{equation}
\lim_{w_i \to \infty} \delta = \delta_{\infty}=-\phi'=-\frac{1}{n\,k_0\,\cos \theta_0}\frac{d \phi}{d \theta}.
\end{equation}

In the asymptotic regime, the second and the third terms of equation \eqref{eq:expansion1} both vanish. The third term vanishes because $\delta^2 \rightarrow (\phi')^2$. The first term is the only one whose limit is not zero but instead

\begin{equation}
\lim_{w_i \to \infty} \Delta = \Delta_{\infty} = \frac{\rho'^2-\rho\,\rho''}{2\rho^2}.\label{e:Delta_inf}
\end{equation}

This very simple formula is the equivalent of Artmann's formula for the width of the beam instead of its position. We underline that only $\rho$ appears in this formula, and that the quantity $\rho'^2-\rho\,\rho''$ plays a central role even outside of the asymptotic limit as shown in expression \eqref{eq:expansion1}.

\begin{figure}
\includegraphics[width=8cm]{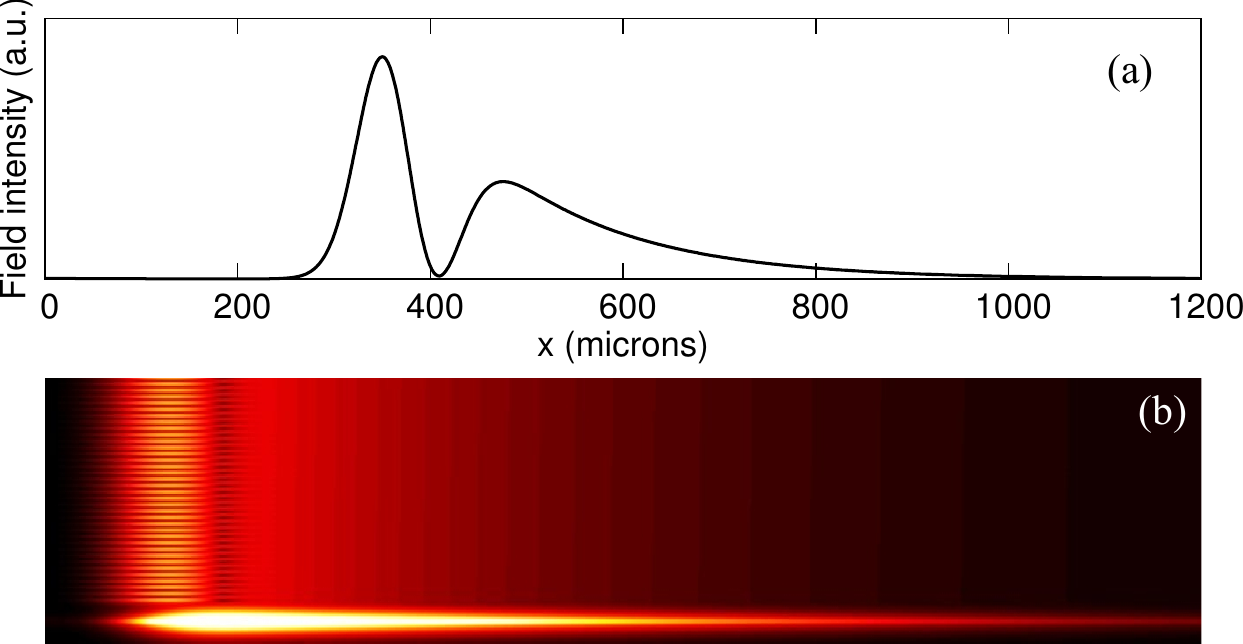}
\caption{Excitation of a leaky mode in a
  waveguide ($\epsilon=5$, thickness of $0.285\lambda$) surrounded by
  air using a gaussian beam (incidence angle $33.9^{\circ}$,
  $w=100\lambda$) propagating in a high index medium (the prism,
  $\epsilon_2=3$). The distance between the prism and the waveguide is
  $0.65\lambda$. (a) Profile of the reflected beam's intensity
  . (b) Map of the corresponding field intensity.
  \label{fig:shift}}
\end{figure}

The formula thus predicts that when $\rho = 1$ whatever the angle, there is simply no change in the reflected beam's width. This may sound correct for total internal reflection, but is quite at odds with conventional knowledge\cite{tamir71a,tamir71b} when the beam is narrow and when a resonance is excited in the structure. For a narrow beam, the excitation of a leaky mode, for instance, is generally expected to lead to a large lateral shift and {\em to a widening} of the reflected beam (see Figure \ref{fig:shift}. Our simulations show that the asymptotic formula is right : in the asymptotic regime when the beam is very large and when $\rho=1$, there is absolutely no change in the width of the beam on reflection, confirming our predictions.

Now, when $\rho$ presents a minimum because of a resonance, then the formula predicts that the reflected beam should be (i) narrower than the incidence beam at resonance because $\rho'=0$ and $\rho''>0$ for a minimum, so that $\rho'^2-\rho\,\rho''<0$ and (ii) wider than the incident beam slightly off resonance when $\rho$ can be considered linear, so that $\rho'$ is maximum and $\rho''$ vanishes which yield $\Delta>0$ for a wide enough beam. And this occurs of course  on an angular range that is much narrower than the dip in the reflection coefficient itself. This leads to think that at resonance precisely, the reflected beam is in general narrower than the incoming beam.

In order to better illustrate this phenomenon and to show its potential, we consider the realistic case of a surface plasmon resonance excited in the Kretschman-Raether configuration at a wavelength of $632.8$ nm, as illustrated on Fig. 1. We have used Moosh\cite{krayzel10,defrance16} to simulate the excitation of the SPR by a gaussian beam ($p$ polarized) propagating in a prism (BK7 glass, with an index of 1.47) with an incidence angle larger than the critical angle of the glass-air interface. A thin gold film ($55$ nm) is attached to the prism with a 2 nm thin chromium layer. These parameters are actually carefully chosen so that the reflection coefficient is not too low at resonance ($\theta_{\rm SPR}=45.5^\circ$,) or the reflected beam would be too weak to allow for any measurement, and to maximize the effect we are looking for. Fig. 1 (b) shows the modulus of the reflection coefficient, $\rho$ as a function of the incidence angle, as well as the quantity $\Delta_{\infty}$. It is obvious how $\Delta_{\infty}$ is supposed to present swift variations. This too is totally at odds with what one would expect for a narrow incoming beam\cite{tamir71a,tamir71b}, since the resonance is the actual excitation of a leaky mode supposed to widen the reflected beam, the surface plasmon.

\begin{figure}
\includegraphics[width=\linewidth]{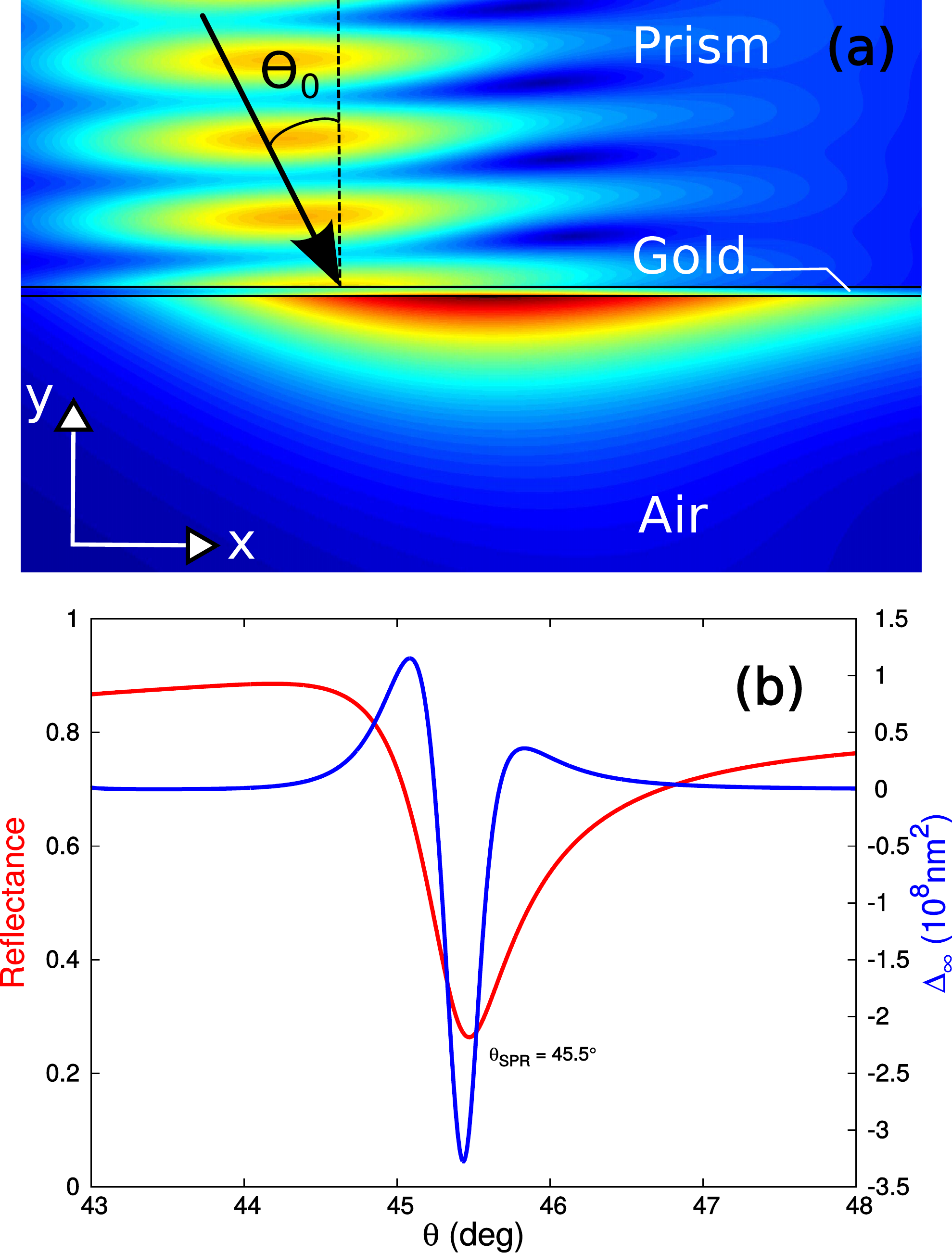}
\caption{(Color online) (a) Modulus of the magnetic field obtained by simulation in the case of a Surface Plasmon Resonance excitation. A gold layer (55 nm) is deposited on the bottom of a prism. The incoming beam comes from above with an incidence angle of $45.5^\circ$ and propagates inside the prism. The reflected beam interferes locally with the incoming beam, hence the fringes. (b) Modulus of the reflection coefficient (red solid line) and asymptotic beam width change (blue solid line).\label{fig:SPR}}
\end{figure}

\begin{figure}[htbp]
\includegraphics[width=\linewidth]{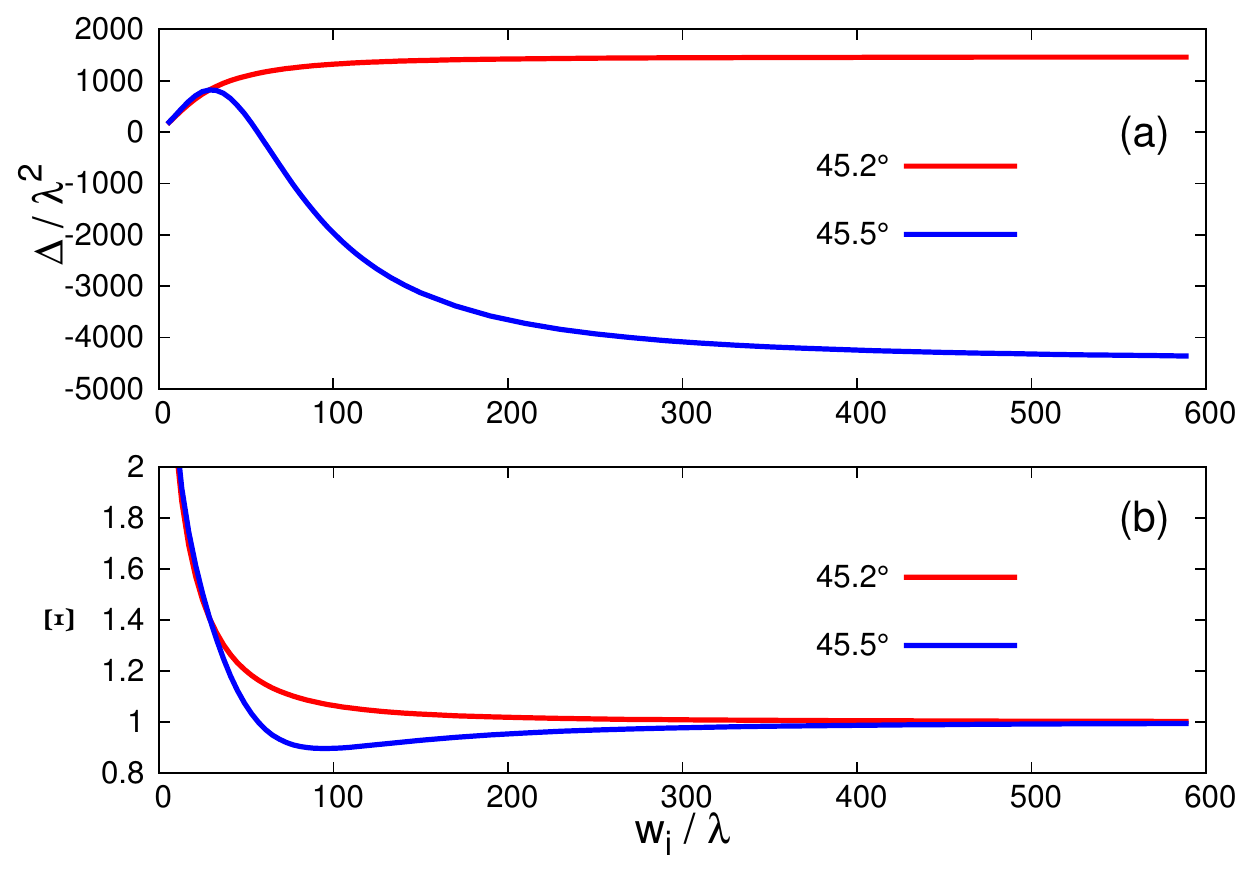}
\caption{(Color online) Absolute ($\Delta$, top) and relative ($\Xi$,bottom) beam width change on reflection for two different incidence angle ($\theta_{\rm SPR}=45.5^\circ$, solid blue line, and $45.2^\circ$, solid red line), as a function of the incoming beam waist (expressed in wavelength units). \label{fig:Delta}}
\end{figure}

We compute $\Delta$ as a function in the incoming beam's width. The results are shown Fig. \ref{fig:Delta} for two different incidence angles. The first angle corresponds exactly to the resonance ($\theta_{\rm SPR}=45.5^\circ$), the other is slightly off resonance ($45.2^\circ$). The absolute widening $\Delta$ is positive off resonance and negative at the precise angle of resonance and the difference between the two behaviour is striking, as shown Fig. 2(a). When the incoming beam is very narrow, no difference can be noticed.

However, the absolute widening is not a perfectly relevant quantity from an experimental point of view. This idea is very important : if a non-specular phenomenon can only be observed for very large beams, then the relative effect will be so small than detecting it can prove impossible.
As the {\em absolute} expansion tends to a limit in the asymptotic regime, the {\em relative} widening defined as the ratio
\begin{equation}
\Xi = \sqrt{\left(\frac{\int (x-\delta)^2 |E_r|^2 dx}{\int |E_r|^2 dx}\right)/\left(\frac{\int x^2 |E_i|^2 dx}{\int |E_i|^2 dx}\right)}
\end{equation}
actually tends to $1$ whatever the angle. This ratio, in the asymptotic limit, is $\frac{w_r^2}{w_i^2}$, but  $w_r$ is generally not well defined since the reflected beam is distorted. This means that there is no relative widening in the asymptotic regime, whereas it is the right quantity to consider if ever we want to measure such a phenomenon experimentally. That is the reason why asymptotic formulae like Artmann's or (\ref{e:Delta_inf}) should not be fully trusted: sometimes the asymptotic regime is so difficult to reach that the relative effect (like the ratio of the lateral shift over the incident beam's waist) is negligible.

Now Fig. 2(b) shows the relative widening as a function of the incoming beam's waist for the two previously chosen incidence angles. As can be seen, both tend to one in the asymptotic regime and are very close when the incoming beam is very narrow, but a clear behaviour difference can still be seen between the two for an intermediate and suprizing low value of $w_i$, well before the asymptotic regime is reached. The difference is actually maximum for $w_i= 75 \lambda$ and represents a $20\%$ relative change in the reflected beam width for a $0.3^\circ$ incidence angle change only. 

This significant change is better illustrated on Fig. 3, where the relative expansion of the reflected beam, $\Xi$, is shown as a function of the incidence angle using the beam width that maximizes this variation. It allows to better capture the very narrow angular range on which the beam width variation occurs. When compared to the change in the reflection coefficient on the same angular range, this leads to think that while the sensitivity of the method would remain the same, monitoring the beam width change would allow to reach a better resolution. 

\begin{figure}[htbp]
\includegraphics[width=\linewidth]{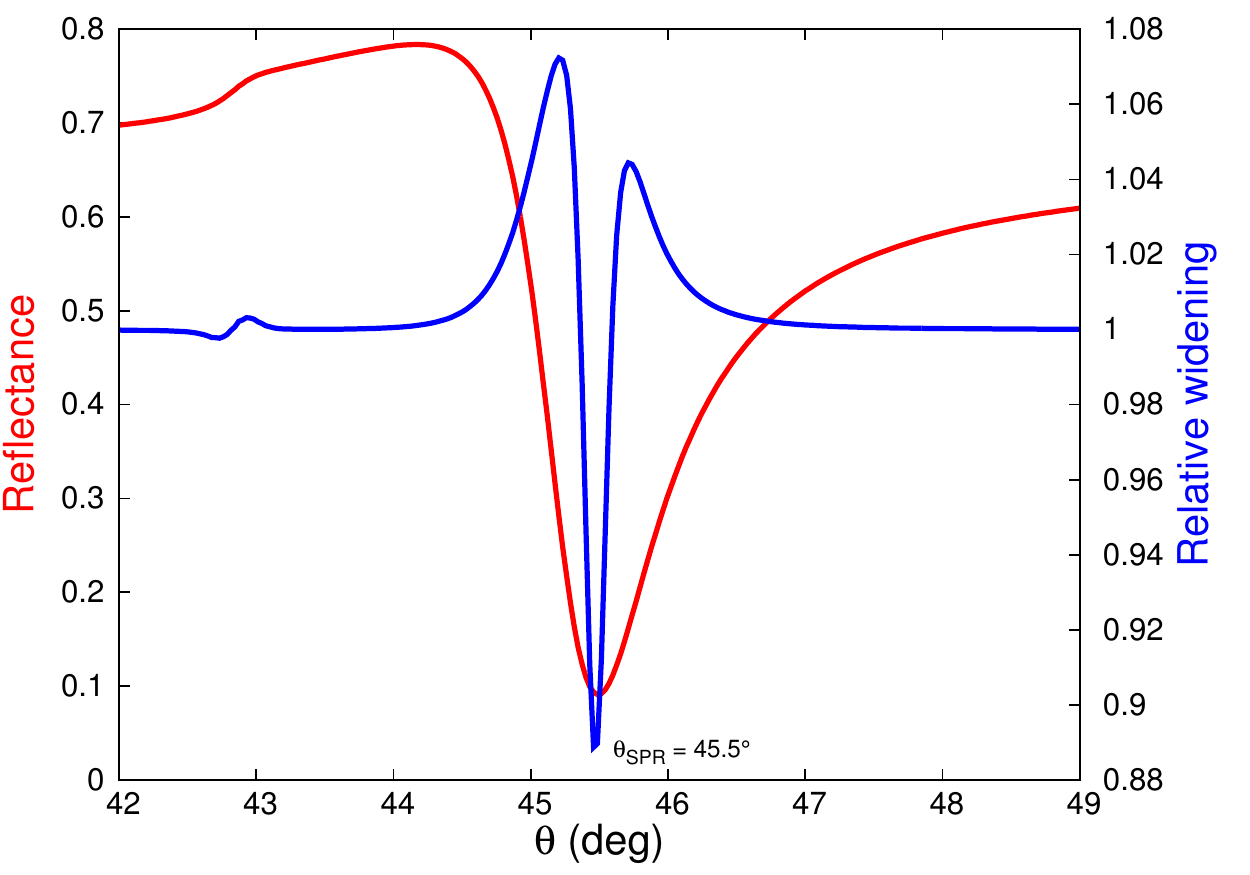}
\caption{(Color online) Reflection coefficient (solid red line) and relative beam with change (solid blue line) on reflection as a function of the incidence angle for an incoming beam with a waist of $75 \lambda$.
\label{fig:3}}
\end{figure}

There are two ways this phenomenon can be physically understood. First, the profile of the reflected beam\cite{profile} can be interpreted as the result of destructive interference between the beam reflected by the first interface between glass and air, and the field leaking out of the surface plasmon itself. At resonance, the interferences are destructive enough to strongly reduce the width of the beam (see Fig. 4 where the profile of the beam at resonance presents a dip, a clear signature of the destructive interference). Sligthly off resonance, the interferences are no longer destructive, so that the leaky mode and the beam reflected by the first interface add up, leading to a widening of the beam. Seen this way, the device can be considered as a new kind of interferometer.

\begin{figure}[htbp]
\includegraphics[width=\linewidth]{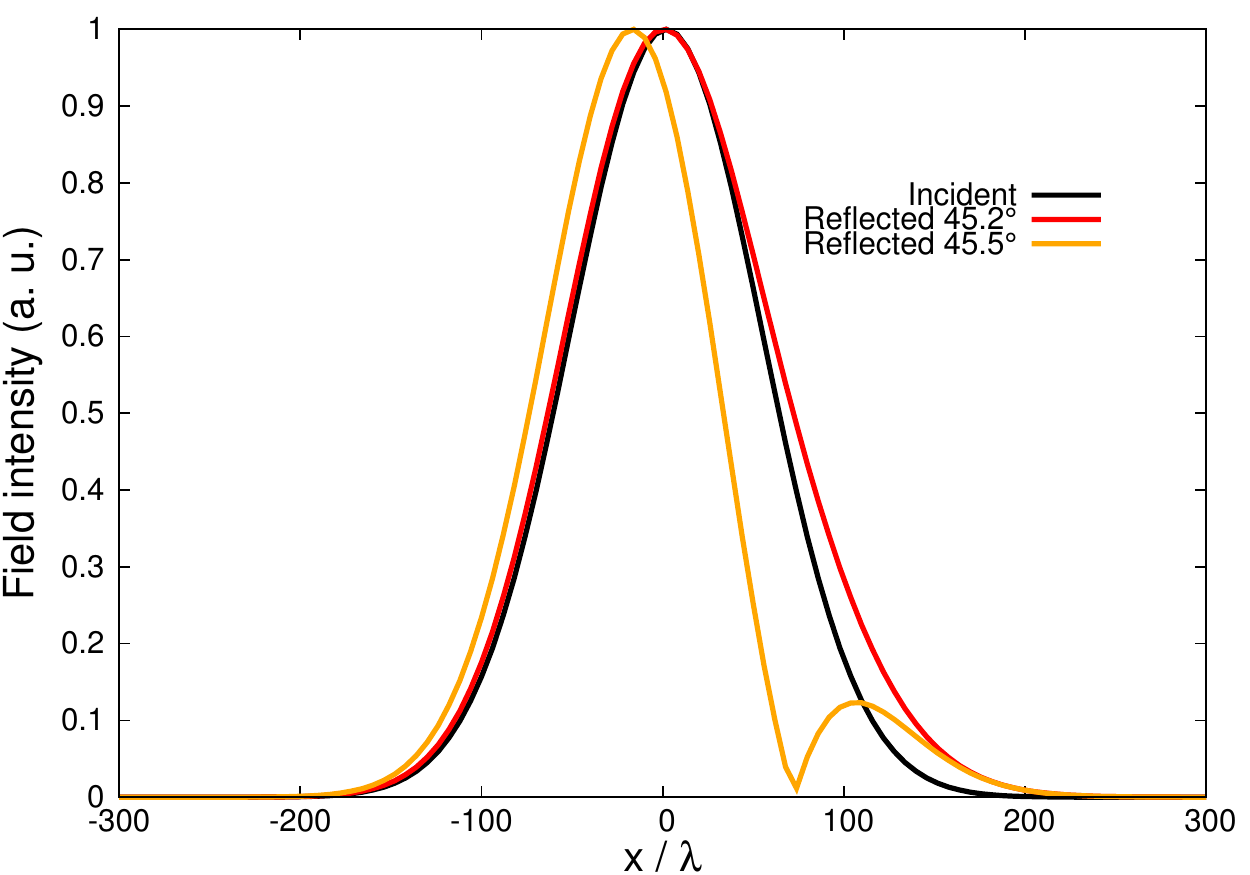}
\caption{(Color online) Profiles of the incoming (solid black line) and of the reflected beam for two incidence angles ($\theta_{SPR}=45.5^\circ$, solid yellow line and $45.2^\circ$ solid red line). The profiles are shown with the same maximum for a better comparison.\label{fig:4}}
\end{figure}

Finally, the whole phenomenon can be understood from a spectral point of view. The angular spectrum of the reflected beam is the angular spectrum of the incident beam times the reflexion coefficient. This allows to understand why the variation in the width of the reflected beam is the largest {\em when the spectral width of the beam is roughly one third of the spectral width of the resonance}. In that case, three domains can be clearly defined, depending on the incidence angle $\theta_0$. For an incidence angle slightly smaller that the resonance angle, $\rho$ is linear and decreasing sharply. The angular spectrum of the reflected beam is thus narrower than the spectrum of the incoming beam and the reflected beam is then spatially larger. The narrowing of the reflected beam occurs when the incoming angular spectrum is centered on the resonance, because the central part of the reflected spectrum is thus diminished, leading to a spectral widening and a spatial narrowing. On the other side of the resonance, the reflected beam is of course spatially widened too.

We have thus provided very general tools to deal with the beam width's change on reflection on a multilayered structure and showed that it is a relevant parameter that can be used to better detect a SPR resonance, eventually opening a new route to improve the resolution of SPR biosensors. This would be the first time since Newton\cite{newton}, that nonspecular phenomenon would find an application.
Furthermore, our work suggests that monitoring non-specular changes {\em outside of the asymptotic regime} is a relevant idea. Asymptotic results are interesting, but as very wide beams are very often required to reach this regime, the effects may be extremely difficult to measure. It is high time, now that we have the numerical tools to deal with realistic finite beams and complex changes in the reflected and transmitted beams, to explore thoroughly what could finally appear as a whole new domain, well beyond the classical nonspecular phenomena as the Goos-H\"anchen or the Imbert-Fedorov lateral shifts. 

We underline that the spectral explanation given above is very general and this resonant narrowing can thus be expected to occur in other domains of physics, like for instance in the case of resonant tunneling in quantum mechanics when a wavepacket is sent on a potential well burried in a barrier\cite{chang74,ng88}. In that case, the part of the wavefunction that is reflected would be, if the conditions are correctly chosen, spatially narrower than the incoming wavefunction despite the time that is spent in the potential well. A physical interpretation is that the beam reflected by the first barrier is interfering destructively with the wavefunction leaking out of the weakly bound state inside the barrier. In electronics, when a stop-band filter is excited with a temporal wavepacket, the resulting signal can be expected to be temporally shorter than the incoming signal in the proper conditions. In all these cases, provided the wavevector $\alpha$ is replaced with the pulsation $\omega$, the formulas that have been given above will correctly describe the resonant narrowing of the temporal wavepacket. 

\section*{Appendix A}

In this first Appendix, we propose a demonstration of Artmann's formula in the asymptotic
regime, showing that the formula is valid even if the modulus of the reflection coefficient
changes with the angle of incidence. Even if the formula has been quite successfully used in that
context, we underline that all the previous demonstration of Artmann's formula have been 
done assuming a reflection coefficient with a unity modulus. The analytical formula that
are necessary to get to the end of the proof will be extremely useful in the following.

The incident and reflected beam fields $E_i$ and $E_r$ can be expressed
whatever the polarization by 
\begin{equation}
E_i(x,z,\omega)=\frac{1}{2\pi}\int \tilde{E_i} \left(\alpha\right) e^{i\left(\alpha x -\gamma z -\omega t \right)}d\alpha,
\end{equation}
and
\begin{equation}
E_r(x,z,\omega)=\frac{1}{2\pi}\int \tilde{E_r} \left(\alpha\right)e^{i\left(\alpha x +\gamma z -\omega t \right)}d\alpha,
\end{equation}
where $\tilde{E_i}$ and $\tilde{E_r}$ are the spectral amplitudes, 
$\gamma=\sqrt{\varepsilon\, \mu\,k_0^2-\alpha^2}$, $k_0$ being the wavenumber in
vacuum and $\varepsilon$ (resp. $\mu$) the permittivity
(resp. permeability) of the upper medium.

The reflection coefficient is defined by
\begin{equation}
r=\rho \, e^{i\phi}=\frac{\tilde{E_r}}{\tilde{E_i}}
\label{eq:reflection}
\end{equation}
where $\rho=\rho(\alpha)$ is the magnitude and $\phi=\phi(\alpha)$ is the phase of $r$.
 
The lateral displacement of the reflected beam is the distance between the
centers of the incident and reflected beams. It can be expressed as
\begin{equation}
\delta=\frac{\int x |E_r|^2 dx}{\int |E_r|^2 dx}-\frac{\int x |E_i|^2 dx}{\int |E_i|^2 dx}.
\end{equation}

Applying the Parseval-Plancherel lemma, we can write:
\begin{equation}
\int x |E_r|^2 dx=\frac{i}{2\pi}\int \frac{\partial \tilde{E_r}}{\partial \alpha} \tilde{E_r}^* d\alpha
\end{equation}
and by inserting expression (\ref{eq:reflection}) we obtain
\begin{align}
\int x |E_r|^2 dx & =\frac{i}{2\pi}\int \frac{\partial }{\partial \alpha}\left( \rho \, e^{i \phi} \tilde{E_i} \right) \rho \, e^{-i \phi} \tilde{E_i}^* d\alpha \\
& = \frac{i}{2\pi}\int \left( \rho \rho' + i \rho^2 \phi' \right) |\tilde{E_i}|^2 d\alpha + \frac{i}{2\pi}\int \rho^2 \frac{\partial \tilde{E_i}}{\partial \alpha} \tilde{E_i}^* d\alpha.
\label{eq:int}
\end{align}
For an incident Gaussian beam the spectral amplitude is
\begin{equation}
\tilde{E_i} \left(\alpha \right) = \frac{w}{2 \sqrt{\pi}} e^{-\frac{w^2}{4} \left(\alpha-\alpha_0 \right)^2} e^{- i\alpha x_0}
\end{equation}
where $x_0$ is the position of the beam's center, given by
$\frac{\int x |E_i|^2 dx}{\int |E_i|^2 dx}$, and
$\alpha_0=\sqrt{\epsilon\,\mu}\, k_0\,sin\,\theta_0$, $\theta_0$ being the angle of incidence of the beam.

In this particular case, we can notice that
\begin{equation}
\frac{\partial \tilde{E_i}}{\partial \alpha} \tilde{E_i}^* = -ix_0 |\tilde{E_i}|^2 - \frac{1}{2} \frac{\partial |\tilde{E_i}|^2}{\partial \alpha},
\end{equation}
so that, using an integration by parts, equation (\ref{eq:int}) yields
\begin{equation}
\int x |E_r|^2 dx = -\frac{1}{2\pi}\int \rho^2 \phi' |\tilde{E_i}|^2 d\alpha + \frac{x_0}{2\pi}\int \rho^2|\tilde{E_i}|^2.
\end{equation}
Finally, the lateral displacement is given by the rigorous formula:
\begin{equation}
\delta=-\frac{\int \rho^2 \phi' |\tilde{E_i}|^2 d\alpha}{\int \rho^2 |\tilde{E_i}|^2 d\alpha}.
\end{equation}
The asymptotic regime is reached when the incident beam is large
enough. The spectral amplitude is then so narrow that $\rho^2$
and $\rho^2 \phi'$ can be considered constant. Another point of view is to
say that the Gaussian function tends towards the Dirac when
$w \rightarrow + \infty$ in the sense of distributions, so that the
asymptotic lateral shift is the same whatever the profile of the beam
\begin{equation}
\lim_{w \to \infty} \delta=-\phi'.
\end{equation}
This result is referred to as Artmann's formula.

\section{Appendix B}

Here we will find an expression for the variation of the reflected beam width. Let us consider a cenetred incident beam for which $\int x|E_i|^2\,dx=0$. The position of the reflected
beam's center, denoted $\delta$, is given by equation (\ref{delta}). The
width of a beam is given by the square root of its second centered momentum, so that the
widening of the reflected beam can be expressed as
\begin{equation}
\Delta=\frac{\int (x-\delta)^2 |E_r|^2 dx}{\int |E_r|^2 dx}-\frac{\int x^2 |E_i|^2 dx}{\int |E_i|^2 dx},
\end{equation}
which can be developed as follows :
\begin{equation}
\Delta=\frac{\int x^2 |E_r|^2 dx}{\int |E_r|^2 dx}-\frac{\int \delta^2 |E_r|^2 dx}{\int |E_r|^2 dx}-\frac{\int x^2 |E_i|^2 dx}{\int |E_i|^2 dx}.
\end{equation}
Applying the Parseval-Plancherel lemma, we get
\begin{equation}
\int x^2 |E_r|^2 dx=-\frac{1}{2\pi}\int \frac{\partial^2 \tilde{E_r}}{\partial \alpha^2} \tilde{E_r}^* d\alpha,
\end{equation}
 and by inserting expression (\ref{eq:reflection}) we obtain
\begin{multline}
2\pi\,\int x^2 |E_r|^2 dx= \\ -\int \left( \rho\rho^{''}+2i\rho\rho'\phi'+i\rho^2\phi''-\rho^2\phi'^2 \right) |\tilde{E_i}|^2 d\alpha \\ - \int \left( \rho\rho'+i\rho^2\phi' \right) \frac{\partial|\tilde{E_i}|^2}{\partial\alpha} d\alpha - \int \rho^2 \tilde{E_i}^*\frac{\partial^2 \tilde{E_i}}{\partial \alpha^2}d\alpha.
\end{multline}
After some integrations by parts, we can write that
\begin{equation}
2\pi\,\int x^2 |E_r|^2 dx=\int(\rho^2\phi'^2-\rho\rho'')|\tilde{E_i}|^2 d\alpha + \int \rho^2 \left(\frac{\partial\tilde{E_i}}{\partial\alpha}\right)^2 d\alpha.
\end{equation}
On an other hand, the equality
\begin{equation}
2\pi\,\int x^2 |E_i|^2 dx=-\int \frac{\partial^2 \tilde{E_i}}{\partial \alpha^2} \tilde{E_i}^* d\alpha
\end{equation}
can be written
\begin{equation}
2\pi\,\int x^2 |E_i|^2 dx=-\int \left(\frac{1}{2}\frac{\partial^2 \left(|\tilde{E_i}|^2\right)}{\partial \alpha^2}-\left(\frac{\partial \tilde{E_i}}{\partial \alpha}\right)^2\right) d\alpha,
\end{equation}
so that we get
\begin{multline}
\Delta = \frac{ \int(\rho^2\phi'^2-\rho\rho'')|\tilde{E_i}|^2 d\alpha } { \int \rho^2|\tilde{E_i}|^2 d\alpha }
+ \frac{ \int \rho^2 \left(\frac{\partial\tilde{E_i}}{\partial\alpha}\right)^2 d\alpha } { \int \rho^2|\tilde{E_i}|^2 d\alpha }\\
+ \frac{ \int \frac{\partial^2 \left(|\tilde{E_i}|^2\right)}{\partial \alpha^2} }{ \int |\tilde{E_i}|^2 d\alpha }
- \frac{ \int \left(\frac{\partial \tilde{E_i}}{\partial \alpha}\right)^2 d\alpha } { \int |\tilde{E_i}|^2 d\alpha }-\frac{\int \delta^2 |E_r|^2 dx}{\int |E_r|^2 dx}.
\label{eq:enlargement}
\end{multline}
For a {\em centered} gaussian beam, equation (\ref{eq:gaussian}) gives
\begin{equation}
\left(\frac{\partial \tilde{E_i}}{\partial \alpha}\right)^2=\frac{w^4}{4}(\alpha-\alpha_0)^2\tilde{E_i}^2
\end{equation}
and
\begin{equation}
\frac{\partial \left(\tilde{E_i}^2\right)}{\partial \alpha}=-w(\alpha-\alpha_0)\tilde{E_i}^2.
\end{equation}
so that
\begin{equation}
\int \rho^2 \left(\frac{\partial\tilde{E_i}}{\partial\alpha}\right)^2 d\alpha = 
-\int \rho^2\frac{w^2}{4}(\alpha-\alpha_0)\frac{\partial \left(\tilde{E_i}^2\right)}{\partial \alpha}d\alpha.
\end{equation}
Using integrations by parts, we obtain
\begin{equation}
\int \rho^2 \left(\frac{\partial\tilde{E_i}}{\partial\alpha}\right)^2 d\alpha =
\int\rho^2\frac{w^2}{4}\tilde{E_i}^2 d\alpha + \frac{1}{2}\int\left(\rho\rho''+\rho'^2\right)\tilde{E_i}^2 d\alpha.
\end{equation}

Following a very similar way, we get 
\begin{equation}
\int \left(\frac{\partial \tilde{E_i}}{\partial \alpha}\right)^2 d\alpha =
\int\frac{w^2}{4}\tilde{E_i}^2 d\alpha.
\end{equation}
Since $\int\frac{1}{2}\partial_\alpha^2
\left(\tilde{E_i}^2\right) \,d\alpha=0$ for any
gaussian (or finite) beam equation (\ref{eq:enlargement}) becomes
\begin{multline}
\Delta = \frac{ \int(\rho^2\phi'^2+\frac{1}{2}(\rho'^2-\rho\rho''))|\tilde{E_i}|^2 d\alpha } { \int \rho^2|\tilde{E_i}|^2 d\alpha }
+ \frac{ \int\rho^2\frac{w^2}{4}\tilde{E_i}^2 d\alpha } { \int \rho^2|\tilde{E_i}|^2 d\alpha }\\
- \frac{ \int\frac{w^2}{4}\tilde{E_i}^2 d\alpha } { \int |\tilde{E_i}|^2 d\alpha }-\frac{\int \delta^2 |E_r|^2 dx}{\int |E_r|^2 dx}.
\label{eq:wide}
\end{multline}

That is the result used in the present paper to estimate the beam width's variation. 

In the asymptotic regime $\delta$ tends towards $-\phi'$ and the terms
two and three of equation (\ref{eq:wide}) cancel each other, so that
\begin{equation}
\lim_{w \to \infty} \Delta=\frac{1}{2}\left(\frac{\rho'^2}{\rho^2}-\frac{\rho''}{\rho}\right).
\end{equation}

Since in the asymptotic limit the reflected beam can be considered as Gaussian, it is relevant to try to link $\Delta$ to a change in the waist of the reflected beam. A straightforward calculation shows that the above formula can in that case be written
\begin{equation}
w_r^2 = w_i^2 +
 2\left(\frac{\rho''}{\rho}-\frac{\rho'^2}{\rho^2}\right)\label{waist-enl}.
\end{equation}

\bibliography{article}

\end{document}